\documentclass[sigconf, nonacm]{acmart}

\usepackage{listings}                                              \usepackage{booktabs}

\definecolor{codegreen}{rgb}{0.13, 0.55, 0.13}
\definecolor{codegray}{rgb}{0.5, 0.5, 0.5}
\definecolor{codeblue}{rgb}{0.0, 0.2, 0.6}
\newcommand\vldbavailabilityurl{https://github.com/SolidLao/GenDB}

\begin{document}
\title{GenDB: The Next Generation of Query Processing — Synthesized, Not Engineered}

\author{Jiale Lao}
\orcid{0009-0003-1144-5152}
\affiliation{%
  \institution{Cornell University}
}
\email{jiale@cs.cornell.edu}

\author{Immanuel Trummer}
\orcid{0000-0002-1579-3221}
\affiliation{%
  \institution{Cornell University}
}
\email{itrummer@cornell.edu}

\begin{abstract}
Traditional query processing relies on engines that are carefully optimized and engineered by many experts. However, new techniques and user requirements evolve rapidly, and existing systems often cannot keep pace. At the same time, these systems are difficult to extend due to their internal complexity, and developing new systems requires substantial engineering effort and cost. In this paper, we argue that recent advances in Large Language Models (LLMs) are starting to shape the next generation of query processing systems.

We propose using LLMs to synthesize execution code for each incoming query, instead of continuously building, extending, and maintaining complex query processing engines. As a proof of concept, we present GenDB, an LLM-powered agentic system that generates instance-optimized and customized query execution code tailored to specific data, workloads, and hardware resources.

We implemented an early prototype of GenDB that uses Claude Code Agent as the underlying component in the multi-agent system, and we evaluate it on OLAP workloads. We use queries from the well-known TPC-H benchmark and also construct a new benchmark designed to reduce potential data leakage from LLM training data. We compare GenDB with state-of-the-art query engines, including DuckDB, Umbra, MonetDB, ClickHouse, and PostgreSQL. GenDB achieves significantly better performance than these systems. Finally, we discuss the current limitations of GenDB and outline future extensions and related research challenges.



\end{abstract}

\maketitle


\vspace{-2em}
\ifdefempty{\vldbavailabilityurl}{}{
\vspace{.3cm}
\begingroup\small\noindent\raggedright\textbf{PVLDB Artifact Availability:}\\
The source code, data, and/or other artifacts have been made available at \url{\vldbavailabilityurl}.
\endgroup
}

\section{Introduction}

A large number of Database Management Systems (DBMSs) have been developed over the decades~\cite{momjian2001postgresql, raasveldt2019duckdb, neumann2020umbra, schulze2024clickhouse, monetdb-hardware-cache-human-better, fernandes2015bigquery, dageville2016snowflake}. These systems feature complex designs and well-engineered implementations to achieve high performance and rich functionality. 
However, new techniques and user requirements evolve rapidly. As a result, existing systems often cannot keep pace and must rely on external extensions or even the development of new systems from scratch. For example, PostgreSQL performs well for OLTP, but it is less effective for OLAP workloads, real-time analytics on time-series data, and workloads that require vector embedding storage and similarity search. Many extensions have been developed to add these capabilities to PostgreSQL~\cite{duckdb_pg_duckdb, timescaledb_github, pgvector_github}, and some extensions attempt to mitigate the architectural constraints of the original system (e.g., the page-oriented storage layout is inefficient for vector indexes~\cite{liu2026fast}). In addition, several purpose-built systems have been designed for these workloads and often achieve significantly higher performance~\cite{shen2023lindorm, wang2021milvus, pinecone_website, neumann2020umbra, schulze2024clickhouse, dageville2016snowflake, fernandes2015bigquery}. 

Due to system complexity, extending database systems is difficult and often introduces incompatibility and system failures~\cite{kim2025anarchy}. Alternatively, developing new systems requires substantial engineering effort and financial cost. Recently, more techniques have emerged, such as semantic query processing over multi-modal data, which has been integrated in some industrial systems~\cite{fernandes2015bigquery, dageville2016snowflake}, and GPU-native analytical processing that has shown better cost efficiency than traditional CPU-based processing~\cite{yogatama2025rethinking}. For these and future techniques, should we continue extending existing systems (e.g., PostgreSQL), or should we design and implement new systems? \textbf{Or is there another option?}


In this paper, we argue that recent advances in LLMs are starting to shape the next generation of query processing: generating query-specific execution code for each query, rather than continuously building, extending, and maintaining complex query engines. As a proof of concept, we present GenDB, an LLM-powered agentic system that generates instance-optimized and customized execution code tailored to specific data, workload, and hardware resources. Generating code per query offers unprecedented space for query optimization and extreme extensibility for integrating new techniques. We present one example below, and discuss more in Section~\ref{sec:research-agenda}.

\begin{example}\label{example:gendb-usecase}
    Query repetition is a key characteristic of real-world workloads: in 50\% of Amazon Redshift clusters, 80\% of queries exactly repeat previously observed queries~\cite{redset-vldb}. Both short- and long-running queries repeat, and long-running queries almost always recur across clusters, as they correspond to regular transformation or analytical tasks~\cite{redset-vldb}. GenDB is well suited for such workloads because the upfront code generation cost can be amortized over many executions. For ad-hoc queries, GenDB can be combined with a traditional DBMS in a hybrid architecture: the traditional system handles one-off queries, while GenDB accelerates frequent queries.
    GenDB serves different users. For developers, it automatically generates instance-optimized execution code, whose correctness can be verified by manual inspection. For users without an SQL background, GenDB can be easily extended with a natural language interface, similar to conversational services deployed in production systems that typically do not provide correctness guarantees~\cite{databricks_bi_2026, google_bigquery_conversational_analytics_2026, snowflake_data_agents_2026}.
\end{example}


GenDB is an agentic system built on LLMs. It takes as input the database schema, natural language questions or SQL queries, and data, along with user configurations (e.g., optimization targets and budgets) and available resources (e.g., hardware resources or access to remote LLM APIs). GenDB first analyzes the workload and hardware to extract structured characteristics that guide subsequent design decisions. It then generates customized storage layouts and index structures. GenDB supports both standard relational data and extended data types, such as images and audio. Based on the workload analysis and storage design, GenDB generates resource-aware execution plans and produces executable code. It uses runtime feedback to iteratively optimize performance. After reaching the user-defined optimization targets or budgets, GenDB outputs the optimized storage structures, indexes, and one executable file per query, all tailored to the specific data, workload, and resources.

We evaluate an early prototype of GenDB on OLAP workloads. We use queries from the TPC-H benchmark and also construct a new benchmark that is designed to reduce potential data leakage from LLM training data. We compare GenDB with state-of-the-art query engines, including DuckDB, Umbra, MonetDB, ClickHouse, and PostgreSQL. By iteratively refining the code based on runtime feedback, GenDB eventually produces correct implementations that achieve significantly better performance. Currently, GenDB validates query correctness by comparing the generated query results with those produced by a traditional database system. When ground truth query results are not available, we recommend that human experts manually verify the generated code. This verification process is generally easier than implementing high-performance code or manually tuning query engines.
For scenarios in which users do not have an SQL background and interact through a natural language interface, GenDB does not provide formal correctness guarantees, which is consistent with conversational services already deployed in production by Databricks~\cite{databricks_bi_2026}, BigQuery~\cite{google_bigquery_conversational_analytics_2026}, and Snowflake~\cite{snowflake_data_agents_2026}. In our future research, we plan to develop additional safety mechanisms, reducing the chances of incorrect code. At the same time, advances in LLMs will help to improve code quality in general. Finally, we discuss the current limitations and outline future extensions and related research challenges.



\section{Background and Related Work}~\label{sec:background}

\noindent \textbf{Compilation-Based Query Code Generation.} The idea of generating customized code for efficient query processing is not new, and GenDB relates to prior work that uses compilation techniques to generate such code~\cite{legobase-query-engine-scala-c, hique-holistic-code-generation-template}. HIQUE~\cite{hique-holistic-code-generation-template} generates customized code for each query by replacing every operator in the execution plan with manually implemented C code templates. The generated code fragments are then compiled and dynamically linked to produce a single executable file. LegoBase~\cite{legobase-query-engine-scala-c} uses a high-level programming language, Scala, to implement query engines, and then employs a compiler to generate optimized C code for each query. These works are motivated by the observation that hand-written C/C++ code clearly outperforms even very fast vectorized systems~\cite{monetdb-hardware-cache-human-better, hyper-compilation-llvm}. This performance gain results from fewer function calls, improved hardware utilization to enhance data and code locality, and more effective compiler optimizations. Hiring human experts to write customized code for a large number of long-running queries is costly and does not scale. Therefore, these methods were proposed as early attempts at automatic code generation. However, these methods provide limited customization because they rely on human-designed code templates and compiler-based optimizations. The generation process itself is fixed and unaware of data characteristics, workload patterns, or hardware properties.

\noindent \textbf{Large Language Model-Augmented Code Generation.} GenDB relates to prior work that uses LLMs for code generation~\cite{codexdb, trummer2025genesisdb, cheng2025letbarbariansinai}. CodexDB~\cite{codexdb} generates Python code for query processing by transforming each operator in an execution plan into a natural language description based on operator-specific text templates, and then invoking the GPT-3 Codex model to generate Python code. GenesisDB~\cite{trummer2025genesisdb} extends CodexDB by generating reusable implementations of general relational operators. These systems provide customization options, such as selecting the data processing library (e.g., Pandas or Vaex in Python) and enabling debugging features (e.g., printing intermediate results during execution). However, they do not target systematic performance optimization and are primarily designed for specific evaluation settings. ADRS~\cite{cheng2025letbarbariansinai} replaces selected algorithms in existing systems with AI-generated alternatives. It iteratively refines prompts, generates or improves solutions using LLMs, and evaluates them in systems. The results show that AI-generated algorithms can outperform human-designed state-of-the-art methods. ADRS is orthogonal to GenDB. ADRS aims to discover effective solutions for existing systems automatically, whereas GenDB replaces traditional engines with synthesized code.

\noindent \textbf{Large Language Model-Enhanced Database Systems.} More broadly, GenDB relates to methods that enhance database systems with LLMs, including DBMS tuning~\cite{gptuner-vldb, gptuner-record, E2ETune, lambda-tune, db-bert}, query rewriting~\cite{song2026quitequeryrewriterules, liu2025genrewritequeryrewritinglarge}, learned query optimization~\cite{llm-query-optimizer, genjoin, rag-llm-query-optimizer}, workload generation~\cite{sqlbarber, sqlbarber-demo, pbench}, semantic query processing~\cite{lao2025sembenchbenchmarksemanticquery, liu2025palimpzest, jo2024thalamusdb, patel2025semanticlotus}, and natural language interfaces~\cite{toxicsql, text2sql-codes, text2sql-evaluation}.

\begin{figure*}[htbp]
    \centering
    \includegraphics[width=0.9\linewidth]{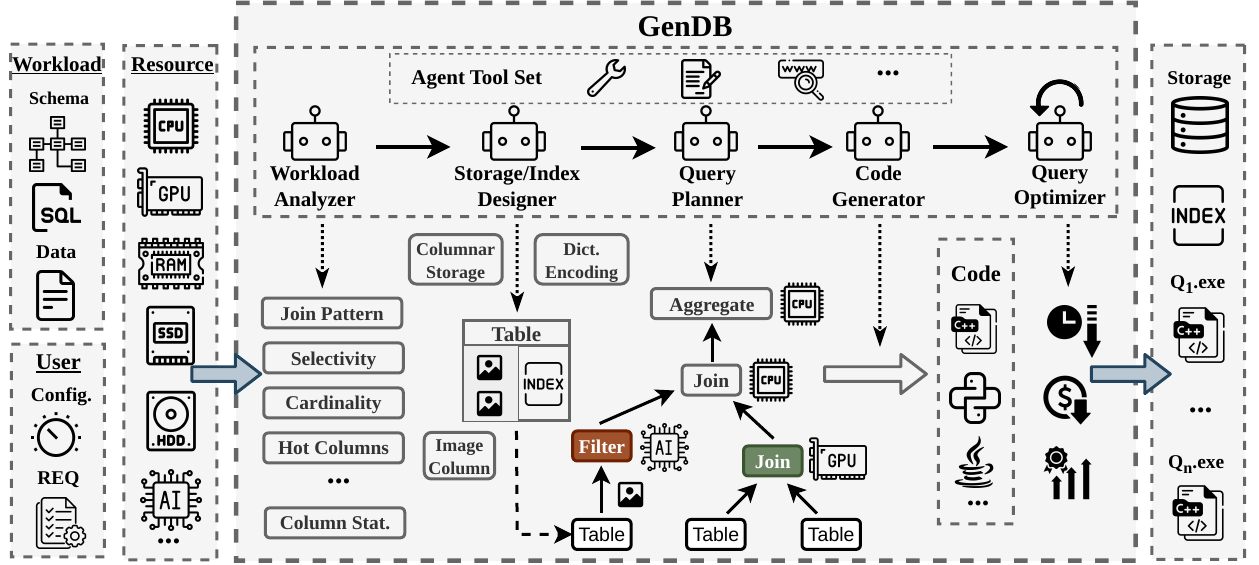}

    \caption{System Overview of GenDB}

    \label{fig:gendb_overview}
\end{figure*}

\section{System Overview}\label{sec:system-overview}

Figure~\ref{fig:gendb_overview} shows an overview of GenDB. It takes as input the target workload, including the database schema, SQL queries or natural language questions, and data; user configurations (e.g., optimization targets and budgets); and available resources (e.g., hardware resources such as GPUs, or access to remote LLMs through APIs for processing multimodal data). Based on these inputs, GenDB generates highly optimized and customized database storage structures (e.g., columnar storage with encoding and compression), indexes, and one executable file per query. All generated components are tailored to the given data, workload, and hardware resources. 

GenDB is an LLM-powered agentic system that decomposes the complex end-to-end query processing and optimization task into a sequence of smaller and well-defined steps, where each step is handled by a dedicated LLM agent.
First, the Workload Analyzer analyzes the hardware resources, database schema, SQL queries, and underlying data to generate structured workload characteristics (e.g., storage type and SIMD support, table and column statistics, join patterns, and filter selectivity). These characteristics are used by downstream agents to guide optimization decisions.
Second, the Storage/Index Designer designs and implements the code that transforms the original data formats into optimized storage structures (e.g., using columnar storage with encoding and compression for OLAP workloads) and builds the corresponding indexes. In the future, we plan to extend GenDB with an extended relational data model in which a cell may store images or audio files in addition to standard SQL data types. Similar models have been adopted in semantic query processing engines~\cite{patel2025semanticlotus, liu2025palimpzest, jo2024thalamusdb, fernandes2015bigquery, dageville2016snowflake}.
Combined with semantic operators powered by LLMs, this extended data model enables the system to process structured and unstructured data in a unified framework. 
Third, the Query Planner generates an efficient execution plan based on the workload analysis and the storage and index designs. Each physical operator implementation is resource-aware. 
For example, the aggregation operator adapts its strategy to the hardware cache hierarchy: for a GROUP BY over a low-cardinality column like nation (25 distinct keys), it uses a direct flat array that fits entirely in L1 cache with zero hashing overhead, whereas for millions of distinct groups it switches to thread-local hash tables sized to fit each core's private cache, merging results in parallel only after the scan completes.
Fourth, the Code Generator implements the execution plan using appropriate programming languages (e.g., C++ for high performance and Python for AI operations) and applies system-level optimizations (e.g., compiler optimizations and operating system functions such as mmap). 
The generated code is executed on the database to measure both correctness and efficiency. 
GenDB validates query correctness by comparing the generated query results with those produced by a traditional database system. If ground truth results are not available, we recommend that human experts verify the generated code, which is typically easier than manually implementing high-performance code or tuning query engines. For users without an SQL background who interact through a natural language interface, GenDB does not provide formal correctness guarantees. This design choice is consistent with existing conversational analytics services in production~\cite{databricks_bi_2026, google_bigquery_conversational_analytics_2026, snowflake_data_agents_2026}.
Fifth, the Query Optimizer uses the feedback to iteratively refine both the execution plan and the code implementation to improve performance. The optimization objective is user-defined and flexible. Currently, we support optimizing a single metric, such as hot run time. Going forward, we plan to experiment with constrained optimization, for example minimizing
memory and disk usage while satisfying a Service Level Agreement (SLA) on query execution time, as well as multi-objective optimization that simultaneously balances performance and cost.
Finally, after exhausting the user-configured optimization budgets (e.g., maximum number of optimization iterations) or reaching the user-specified optimization targets, GenDB outputs the instance-optimized data storage, indexes, and one executable file per query, all tailored to the specific data, workload, and available resources.

Each agent in GenDB uses an LLM as its core reasoning component to plan and automatically invoke tools to complete the assigned tasks. Each agent is equipped with built-in tools, including file operations such as reading, writing, and editing files; terminal access for executing commands and scripts; and web-related tools such as web search and web content retrieval. The agent capability can be further extended by integrating Agent Skills~\cite{li2026skillsbenchbenchmarkingagentskills} or by implementing the Model Context Protocol (MCP)~\cite{ehtesham2025surveyagentinteroperabilityprotocols}. This tool-use capability allows agents to complete complex tasks. For example, if a query contains multi-table joins and exhibits very poor performance, an agent can generate a sampling program to evaluate candidate join orders, measure their performance, and then refine the code implementation to use the empirically best join order.



\begin{figure*}[htbp]
    \centering
    \includegraphics[width=0.9\linewidth]{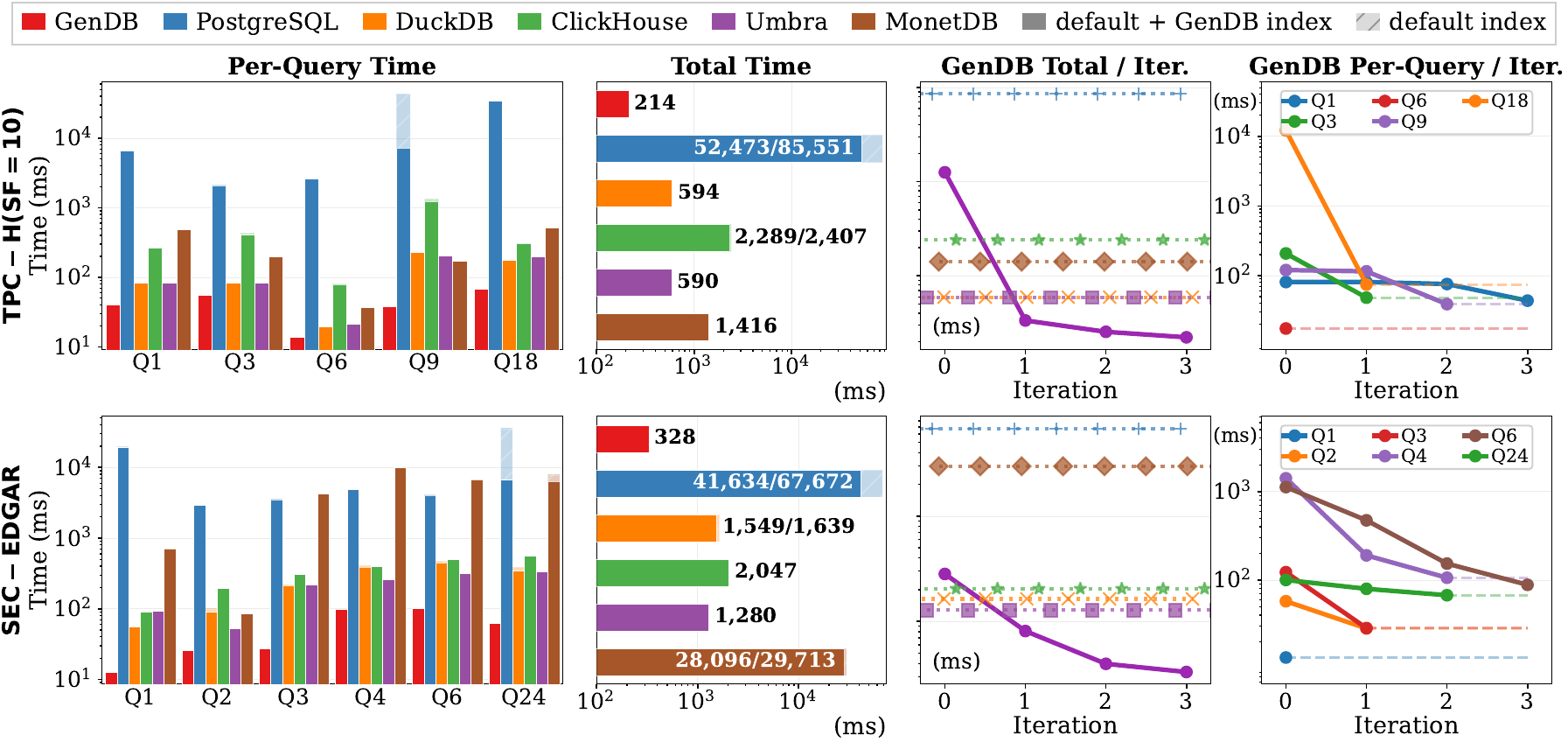}

    \caption{Experimental Results on TPC-H (SF = 10) and SEC-EDGAR Benchmarks}

    \label{fig:exp_results}
\end{figure*}

\section{Experiments}~\label{sec:experiments}

\vspace{-1em}
\subsection{Setup}\label{subsec:setup}

\noindent \textbf{Compared Systems.} 
We compare GenDB with state-of-the-art query engines, using their latest versions at the time of writing: DuckDB v1.4.4~\cite{raasveldt2019duckdb}, ClickHouse v26.2.1~\cite{schulze2024clickhouse}, Umbra (February 2026 release)~\cite{neumann2020umbra}, MonetDB v11.55.1~\cite{idreos2012monetdb}, and PostgreSQL v18.2~\cite{momjian2001postgresql}. 
We ensure that all systems use comparable hardware resources.
GenDB does not require manual tuning of system parameters or indexes, because it determines suitable settings during the code generation process. 
To ensure fairness, we report two versions: one where each system uses only its automatically created indexes (e.g., primary key indexes, or zonemaps that OLAP engines create by default), and one where these system-created indexes are supplemented with additional indexes generated by GenDB.

\noindent \textbf{Hardware.} All experiments are conducted on a Ubuntu server with two Intel Xeon Gold 5218 CPUs (2.3GHz with 32 physical cores).
The system has 384 GB of RAM, which ensures that the entire database is cached in memory, following the same configuration as in~\cite{leis2015good}. We report the execution time of hot runs for each system.

\noindent \textbf{Implementation.} GenDB is implemented in JavaScript and employs Claude Agent to build the agentic system. Claude Sonnet 4.6 serves as the underlying LLM, selected to balance cost and performance. The generated code is implemented in C++.

\noindent \textbf{Benchmarks.} We use (1) TPC-H~\cite{TPC2013} with scale factor 10, and (2) a new manually constructed benchmark, {SEC-EDGAR}.
We report detailed results for individual queries rather than only aggregate results. Following~\cite{compile-vectorize}, we select a representative subset (Q1, Q3, Q6, Q9, Q18), which captures the main performance challenges in TPC-H. Prior analysis shows that an engine that performs well on these queries is likely to perform well on the full workload~\cite{tpc-h-analyzed}.
We introduce SEC-EDGAR to reduce potential data leakage. TPC-H is well studied, and its queries and optimization strategies are well-represented in LLM training data. In contrast, SEC-EDGAR is based on a real-world financial statement dataset that has rarely been used for database benchmarking. We use data from 2022 to 2024 (5~GB)~\cite{SEC_AQFS}.  We generate 1,000 random queries using SQLSmith~\cite{sqlsmith} and then apply diversity-based sampling to select six queries.

\subsection{Experimental Results}\label{subsec:exp}

Figure~\ref{fig:exp_results} shows the experimental results, and GenDB outperforms all baselines on every query in both benchmarks. On TPC-H, GenDB achieves a total execution time of \textbf{214 ms} across five representative queries. This result is \textbf{2.8$\times$} faster than DuckDB (594 ms) and Umbra (590 ms), which are the two fastest baselines, and \textbf{11.2$\times$} faster than ClickHouse. On SEC-EDGAR, GenDB achieves \textbf{328 ms}, which is \textbf{5.0$\times$} faster than DuckDB and \textbf{3.9$\times$} faster than Umbra. 
The performance gap increases with query complexity. For example, on TPC-H Q9, which is a five-way join with a \texttt{LIKE} filter, GenDB completes in 38~ms, which is \textbf{6.1$\times$} faster than DuckDB. 

GenDB uses iterative optimization with early stopping criteria. On TPC-H, Q6 reaches a near-optimal time of 17 ms at iteration~0 with zone-map pruning and a branchless scan, and does not require further optimization. In contrast, Q18 starts at 12,147~ms and decreases to 74~ms by iteration~1, which is a \textbf{163$\times$} improvement. This gain comes from replacing a cache-thrashing hash aggregation with an index-aware sequential scan. 
On SEC-EDGAR, Q4 decreases from 1,410~ms to 106~ms over three iterations, which is a \textbf{13.3$\times$} improvement, and Q6 decreases from 1,121~ms to 88~ms over four iterations, which is a \textbf{12.7$\times$} improvement. In Q6, the optimizer gradually fuses scan, compact, and merge operations into a single OpenMP parallel region, which removes three thread-spawn overheads.
By iteration~1, GenDB already outperforms all baselines.

GenDB achieves high performance through workload-specific and hardware-aware optimizations that are difficult to implement in general-purpose engines. We group these into four categories. First, \emph{data-aware column encoding}: GenDB analyzes value distributions to select compact physical layouts that reduce memory traffic. Second, \emph{algorithm-level restructuring}: rather than selecting from a fixed set of join and aggregation operators, GenDB generates algorithms tailored to the data layout and available indexes. Third, \emph{cache-adaptive aggregation}: GenDB chooses aggregation strategies based on group cardinality and cache capacity, for example using direct arrays for low-cardinality groups and hash tables otherwise. Fourth, \emph{workload-specific derived data structures}: GenDB builds query-specific structures such as nibble-packed lookup files and batched-prefetch probe patterns. These optimizations require joint reasoning across storage layout, query semantics, data distributions, and hardware characteristics, which fixed operator abstractions in general-purpose engines cannot provide.

Due to space limitations, we present one representative example. A general-purpose engine uses the same hash aggregation operator for all queries. In contrast, GenDB generates different aggregation code based on the number of groups produced by the query and the cache level in which the aggregation state resides, as illustrated in the code snippets below. The test machine has 64 cores, each with 32\,KB L1 and 512\,KB L2, sharing a 44\,MB L3.
TPC-H Q1 groups by \texttt{returnflag} $\times$ \texttt{linestatus}, which produces only $3 \times 2 = 6$ groups. Each accumulator holds 6 fields (\texttt{cnt}, \texttt{sum\_qty}, \texttt{sum\_price}, etc.) and is padded to a 64-byte cache line to prevent false sharing, so 6 accumulators occupy only $6 \times 64 = 384$ bytes per thread, which fits entirely in the 32\,KB L1. GenDB therefore skips hashing and uses a direct array. The group index is computed by a simple multiply-add on the two column values. After all 64 threads finish, merging is a sequential sum over $6 \times 64 = 384$ values.
TPC-H Q3 groups by order key with up to 4M distinct groups. If each of the 64 threads kept its own 4M-entry hash table, the total memory would be $64 \times 4\text{M} \times 12\text{B} \approx 3$\,GB (each entry stores a 4-byte key and an 8-byte revenue value), far exceeding the 44\,MB L3. GenDB therefore uses a single shared hash table with a column-separated layout: the keys (\texttt{int32\_t}, $4\text{M} \times 4\text{B} = 16$\,MB) are stored separately from the values (\texttt{double}, $4\text{M} \times 8\text{B} = 32$\,MB), so that probes only touch the key array, which fits in the 44\,MB L3. Multiple threads update the shared value array via lock-free compare-and-swap (CAS), which avoids both locks and a separate merge phase. Between these two extremes, GenDB similarly sizes per-thread hash tables to fit in L2 or L3 depending on the group count, and selects the appropriate merge strategy accordingly. This adaptation requires knowing both the group cardinality and cache topology at code generation time, which general-purpose engines cannot do because they must handle arbitrary cardinalities with a single operator implementation.
\begin{lstlisting}[title={\footnotesize\textbf{TPC-H Q1 (6 groups):} Direct array in L1, no hashing},label={lst:q1-agg}]
// 6 groups x 64 B = 384 B/thread, fits in 32 KB L1
struct alignas(64) Accum {int64_t cnt, sum_qty, sum_price, ...;};
std::array<Accum, 6> local; // per-thread, no hashing
int g = returnflag[i] * 2 + linestatus[i]; // direct index
local[g].cnt++;
local[g].sum_price += price[i];
\end{lstlisting}

\begin{lstlisting}[title={\footnotesize\textbf{TPC-H Q3 (4M groups):} Shared column-layout hash table, keys in L3}]
// Column-separated: keys and values in separate arrays
int32_t oa_keys[4194304]; // 4M x 4B = 16 MB, fits in 44 MB L3
double  oa_rev[4194304];  // 4M x 8B = 32 MB

// Probe only touches oa_keys (L3 hit), not oa_rev
uint64_t slot = oa_probe(oa_keys, orderkey);
// 64 threads share one table; lock-free CAS, no merge needed
atomic_add_double(&oa_rev[slot], revenue);
\end{lstlisting}\label{code-Q2}

\begin{figure}[htbp]
    \centering
    \includegraphics[width=0.9\linewidth]{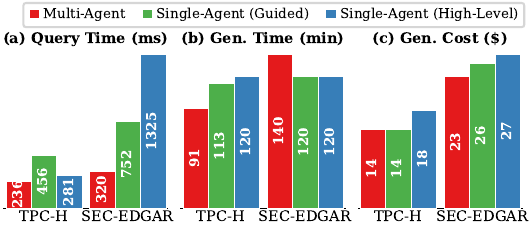}

    \caption{Ablation Study of GenDB}
    \label{fig:ablation-study}
\end{figure}

Figure~\ref{fig:ablation-study} presents an ablation study of GenDB. Starting from the full multi-agent system, we simplify it in two steps. First, we replace the multi-agent pipeline with a single agent that still receives domain-specific hints about database optimization strategies (the \emph{guided} variant). Second, we remove these hints entirely and give the single agent full autonomy to decide its own approach (the \emph{high-level} variant).
On TPC-H, the multi-agent design achieves 236~ms, which is \textbf{1.2$\times$} faster than the best single-agent variant (guided, 281~ms). On the unseen SEC-EDGAR benchmark, the gap widens to \textbf{2.3$\times$} over guided (752~ms) and \textbf{4.1$\times$} over high-level (1,325~ms). The multi-agent design also costs less: \$14.15 vs.\ \$17.54 on TPC-H and \$23.49 vs.\ \$27.46 on SEC-EDGAR.
Both ablation steps show a clear effect. On TPC-H, the single-agent variants remain competitive because its queries and optimization strategies are well-represented in LLM training data, but they degrade on the unseen SEC-EDGAR benchmark. Removing domain-specific hints (guided $\to$ high-level) causes a \textbf{1.8$\times$} slowdown on SEC-EDGAR, showing that even coarse task guidance helps LLMs generalize beyond memorized patterns. Restoring multi-agent coordination (guided $\to$ multi-agent) improves performance further: the pipeline is decomposed into well-defined steps---analysis, storage design, planning, code generation, and optimization---each handled by a specialized agent. The multi-agent design also uses deterministic code to enforce constraints (e.g., iteration budgets, stopping criteria) and structured JSON for inter-agent communication, whereas the single agent may spawn sub-agents that omit user-specified constraints and rely on less structured free-form text exchanges.


\section{Research Agenda}~\label{sec:research-agenda}


\subsection{Reducing Code Generation Cost}\label{subsec:reduce-cost}

\noindent \textbf{Selecting Models Adaptively Based on Task Complexity.} Using smaller models can reduce both generation time and monetary cost, but may also reduce output quality. A promising direction is to assign models of different capability at two levels. First, across pipeline steps: our ablation study shows that structured decomposition improves both quality and cost, suggesting that simpler steps such as analysis could use smaller models while complex steps such as code generation require larger ones. Second, across queries: TPC-H Q6 converges at iteration~0 while Q18 requires a \textbf{163$\times$} improvement over multiple iterations, indicating that we can use different models and efforts for queries of different complexities.

\noindent \textbf{Generating Reusable Components.} Instead of generating code for each query independently, GenDB can reduce total generation cost by generating reusable operators shared across queries or by generating code for query templates. Currently, GenDB only reuses basic utility functions (e.g., date transformation). A promising direction is to generate reusable data structures, relational operators, and query templates. Our experiments suggest this is feasible: cache-adaptive aggregation applies to multiple TPC-H queries, differing only in group cardinality, and such patterns could be captured as parameterized components. Beyond sharing, multi-query optimization~\cite{multi-query-optimization} could allow GenDB to produce a single executable that returns results for multiple queries in one execution.

\noindent \textbf{Reducing Token Consumption.} Token consumption can be reduced by designing more concise prompt templates, configuring LLMs with lower reasoning effort, and applying compression methods to inter-agent communication (e.g., TOON~\cite{masciari2026llms}), database schemas~\cite{schema-compression}, and queries~\cite{lambda-tune}. However, these approaches may degrade generation quality. It remains an open challenge to balance the degree of token consumption and generation quality.

\subsection{Improving Code Generation Robustness}\label{subsec:improve-robust}

\noindent \textbf{Integrating Domain Knowledge into Code Generation.} 
Currently, GenDB does not use an external knowledge base, as most domain knowledge is already contained in LLM training data. However, even when relevant knowledge exists in the training data, an LLM may retrieve it incorrectly or inconsistently from its parameters. More importantly, non-public knowledge---such as unpublished research, proprietary optimization techniques, and company-internal practices---is entirely absent from the training data yet could substantially improve generation quality. Public knowledge from recent work can in principle be retrieved via web search, but the web is a massive and noisy source, making effective and efficient retrieval a challenge in itself. A promising direction is to integrate and manage an external knowledge base that supplements the LLM's parametric knowledge, while preventing GenDB from being misled by noisy or unverified information.


\noindent \textbf{Handling Silent Agent and LLM Failures.} We identify two categories of silent failures that occur during code generation. First, LLM-generated code can be inefficient or non-terminating, especially for complex schemas and queries. A common cause is incorrect hash table construction for joins and aggregations with open addressing: LLMs may omit resizing logic or probe bounds, underestimate cardinality, or let the load factor approach 100\%, resulting in non-terminating probe loops. We observed this particularly on queries with high-cardinality joins such as TPC-H Q18, where the optimizer must reason carefully about table sizing across 64 threads. Second, when the task is complex, the total number of reasoning and output tokens may exceed the LLM's maximum output limit, causing the agent to hang without returning a response~\cite{claude_issue_24055}. Both failure types can cause the generation process to stall indefinitely and waste server resources. Our current solution enforces a query-level timeout (300 seconds) and an agent-level timeout (30 minutes), automatically rolling back to the best implementation from previous iterations if either is triggered. More systematic solutions remain for future work, such as automatic detection of potential code issues with graceful termination and explicit error reporting.

\noindent \textbf{Improving and Scaling Inter-Agent Communication.} Inter-agent misalignment is a major cause of failure~\cite{pan2025multiagent-fail}, and is particularly pronounced in code generation for query processing, where message complexity grows with schema size and query complexity. Our ablation study confirms that structured communication matters: the multi-agent design uses structured JSON for inter-agent messages, whereas the single agent---which may still spawn sub-agents and thus also relies on inter-agent communication---uses less structured free-form text, contributing to its weaker performance on unseen workloads.
We identified one issue when the Storage/Index Designer represents \texttt{DECIMAL} columns as \texttt{int64\_t} with a \texttt{scale\_factor}, recorded in a per-query document for downstream agents. However, later agents do not always follow this specification--for instance, omitting the required division by the factor after multiplying two \texttt{DECIMAL} columns, which produces incorrect results. Currently, GenDB uses JSON and predefined Markdown templates. Evaluating additional protocols such as TOON~\cite{masciari2026llms}, MCP~\cite{anthropic2024mcp}, and A2A~\cite{ehtesham2025surveyagentinteroperabilityprotocols} is a promising direction for future work.

\noindent \textbf{Evaluating and Verifying Generated Code.} 
GenDB relies on feedback to iteratively correct and optimize query implementations. Currently, we validate correctness by comparing GenDB's output against ground-truth results from a traditional DBMS (e.g., DuckDB), and ensure reliable measurement through a notification-based execution queue that serializes query execution while parallelizing planning and code generation across queries. 
Going forward, we plan to explore three directions. First, increasing execution throughput while preserving measurement accuracy, since the current serialized queue is a bottleneck. Second, stronger correctness guarantees such as formal verification for templated query patterns. Third, more fine-grained feedback, such as per-operator profiling or hardware counter analysis, to help the optimizer identify bottlenecks more precisely and converge in fewer iterations.


\noindent \textbf{Developing a Self-Evolving System that Improves over Time.} Generating instance-optimized execution code is an iterative process that often requires trial and error. Instead of running GenDB repeatedly and relying on humans to summarize observations to refine the system, we should design a self-evolving system that learns from past trials and improves over time.
One possible extension is to maintain a persistent module called ``experience.'' After each generation task, an agent could analyze the run and summarize (1) techniques that improve performance and (2) issues that lead to incorrect query results or inefficient execution. The system could then integrate validated rules into the prompts that guide the LLMs, or redesign the system according to the accumulated experience.

\subsection{Extending Code Generation Flexibility}



\noindent \textbf{Generating Customized Code for Semantic Query Processing.} Database systems are increasingly extended to support diverse data types such as text, images, audio, and video~\cite{madden2024databases, lao2025sembenchbenchmarksemanticquery, fernandes2015bigquery, dageville2016snowflake, liu2025palimpzest, patel2025semanticlotus, jo2024thalamusdb}, often using ML models or LLMs to process multimodal data (e.g., {AI.IF(picture, ``the picture shows a red car'')}). In such workloads, model inference dominates execution cost, shifting the optimization focus from relational operators to reducing inference cost and improving result quality. GenDB naturally extends to this setting: it can generate Python code that directly deploys ML models or invokes remote LLMs, and explore techniques such as prompt tuning and model selection~\cite{lao2025sembenchbenchmarksemanticquery} to optimize cost and quality.

\noindent \textbf{Generating Specialized Code for Modern Hardware.} 
GPU-powered analytical processing has shown strong performance and, in many cases, surpasses traditional CPU-based processing in cost efficiency~\cite{yogatama2025rethinking, li2025scaling}. In addition, several high-performance GPU libraries can be leveraged by GenDB to generate GPU-native query processing code. For example, libcudf~\cite{rapids_cudf} provides GPU-efficient implementations of relational operators, and RMM~\cite{rapids_rmm} supports high-performance GPU memory management.

\section{Conclusion}\label{sec:conclusion}

GenDB shifts query processing from manually engineered systems to LLM-driven per-query code generation. Our prototype shows promising results while indicating room for improvement.



\bibliographystyle{ACM-Reference-Format}
\bibliography{sample}

\end{document}